\documentclass[a4paper,12pt]{article}
\usepackage{microtype}
\usepackage[natbibapa]{apacite}
\usepackage{endnotes}
\usepackage[english]{babel}
\usepackage[utf8x]{inputenc}
\usepackage{amsmath}
\usepackage{graphicx}
\usepackage{tabularx}
\usepackage{footmisc}
\usepackage{setspace}
\usepackage{booktabs}
\usepackage[hyperindex,breaklinks]{hyperref}

\usepackage{breakurl}
\usepackage[usenames,dvipsnames,svgnames,table]{xcolor}
\usepackage{a4wide}
\usepackage{enumitem}
\usepackage[normalem]{ulem}
\usepackage{times}
\usepackage{xr}
\usepackage{acronym}
\acrodef{rq:clusters}[\ref{rq:clusters}]{Are there common patterns in how entities emerge in online text streams?} 
\acrodef{rq:substreams}[\ref{rq:substreams}]{Do news and social media text streams exhibit different emergence patterns?} 
\acrodef{rq:entities}[\ref{rq:entities}]{Do different types of entities exhibit different emergence patterns?} 
\acrodef{CM}{collective memory}
\acrodef{EB}{early bursting}
\acrodef{LB}{late bursting}

\title{The Birth of Collective Memories: Analyzing Emerging Entities in Text Streams}

\author{
David Graus\textsuperscript{1,2}\thanks{Corresponding author.}\qquad
Daan Odijk\textsuperscript{3}\qquad
Maarten de Rijke\textsuperscript{1}
\\[1.2ex]
\textsuperscript{1}~Informatics Institute, University of Amsterdam\\ Amsterdam, The Netherlands\\
\textsuperscript{2}~FD Mediagroep, Amsterdam, The Netherlands\\
\textsuperscript{3}~Blendle, Utrecht, The Netherlands
\\[1.2ex]
dpgraus@gmail.com, daan@blendle.com, derijke@uva.nl
}

\date{}

\begin{document}
\maketitle

\begin{abstract}
We study how collective memories are formed online. We do so by tracking entities that emerge in public discourse, that is, in online text streams such as social media and news streams, before they are incorporated into Wikipedia, which, we argue, can be viewed as an online place for collective memory. By tracking how entities emerge in public discourse, i.e., the temporal patterns between their first mention in online text streams and subsequent incorporation into collective memory, we gain insights into how the collective remembrance process happens online. Specifically, we analyze nearly 80,000 entities as they emerge in online text streams before they are incorporated into Wikipedia. The online text streams we use for our analysis comprise of social media and news streams, and span over 579 million documents in a timespan of 18 months. 

We discover two main emergence patterns: entities that emerge in a ``bursty'' fashion, i.e., that appear in public discourse without a precedent, blast into activity and transition into collective memory. Other entities display a ``delayed'' pattern, where they appear in public discourse, experience a period of inactivity, and then resurface before transitioning into our cultural collective memory. 
\end{abstract}


\section*{Introduction}
\label{sec:introduction}

Remembering is a social process~\cite[]{halbwachs1997memoire}. Collective remembrance is the process in which information moves from public discourse into a shared collective memory. 
This process has been compared to the remembrance process of an individual, whose memories transfer from short-term into long-term memory~\cite[]{assmann1995}. 
This comparison has been formalized by mapping the collective's equivalent of long-term and short-term memory to the \emph{cultural} and \emph{communicative} memory, respectively. 

\emph{Cultural} \ac{CM}, the collective's equivalent of an individual's long-term memory, is characterized by being organized, specialized, formal, structured, and distanced from the immediate~\cite[]{assmann1995}. 
Wikipedia is known to ``democratize information,'' through its collaborative nature: its content is produced by volunteer editors and authors from around the world~\cite[]{10.2307/20864471}. Wikipedia has been called an online place for cultural \ac{CM}~\cite[]{Pentzold01052009,ASI:ASI23518}. 
We support this view, and argue that the aforementioned characteristics fit Wikipedia's nature. 
First, Wikipedia is \emph{organized}, through its hierarchy of contributors, where authors are distinguished from admins. 
Wikipedia is \emph{specialized}, since appropriately citing relevant and expert sources to support and back up newly added information is a requirement.
These conventions, requirements, and policies around contributing new information to Wikipedia impose a level of \emph{formality} and enable its coherent and consistent \emph{structure}. 
Finally, the requirement for new articles to be collectively deemed ``important enough,'' ensures Wikipedia's \emph{distance} from the immediate. 

\emph{Communicative} \ac{CM} is in many aspects the opposite of cultural \ac{CM}. 
Analogously to an individual's short-term memory, communicative \ac{CM} is mainly orally negotiated, close to the everyday, disorganized, informal, and non-specialized~\cite[]{assmann1995}. 
Online text streams fit this notion of orally negotiated memory: 
the rapid pace and high volume at which content is published by news websites and social media platforms means that---as opposed to the carefully curated and edited nature of Wikipedia---online text streams are close to the everyday: they not only record and reflect the actions of everyday life but also have a role in producing everyday life for a media-enabled public~\cite[][p.~33]{tierney2013public}. 
With the advent of Web 2.0, and the ability for anyone to publish content on the web, online text streams have naturally become disorganized, informal, and non-specialized. 

We study the evolution of collective memory by tracking additions to our online cultural \ac{CM}, Wikipedia. 
Specifically, we study real world \emph{entities}\footnote{By an ``entity'' we follow standard practice and mean a \emph{thing with distinct and independent existence}, e.g., a person or device (Oxford Dictionary, \url{http://www.oxforddictionaries.com/definition/english/entity}).} as they emerge in online text streams, and are subsequently added to Wikipedia. 
Every day, new content is being added to Wikipedia, with the knowledge base receiving over 6 million monthly edits at its peak~\cite[]{Suh:2009:SNS:1641309.1641322}. 
Domain experts may find information missing on Wikipedia and take up the task of contributing this new information. 
Alternatively, new, previously unheard-of  entities may emerge in news articles or social media postings that describe or comment on events, e.g., 
the Olympics may introduce new athletes onto the world stage, 
or the opening of a new restaurant may be reported in local news media and appear in social media. 
Studying entities that transition from public discourse into Wikipedia gives us insights into how collective memory evolves---for the first time, the online world allows us to make such observations at scale. 

\begin{figure}[t]
\centering
\includegraphics[width=.95\textwidth]{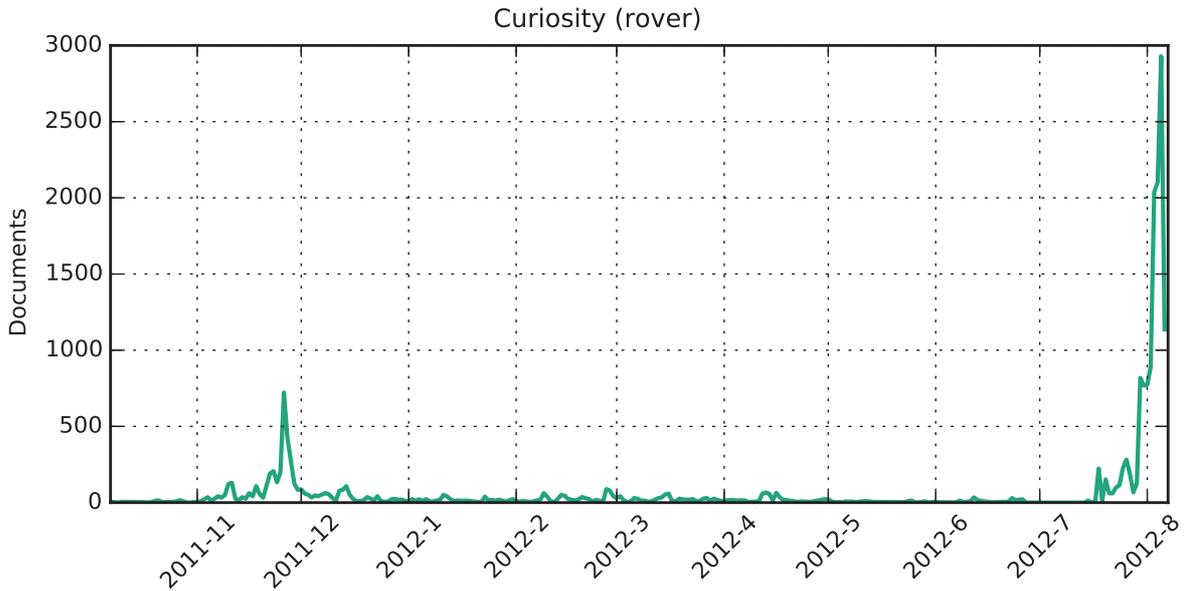}
\caption{Emergence pattern of the entity \texttt{Curiosity (Rover)}, first mentioned in our text stream in October 2011. 
The Wikipedia page for Curiosity was created nine months later, on August 6, 2012. 
There are two distinct \emph{bursts}, one late November 2011, the second shortly before the entity is added to Wikipedia. 
The two bursts correspond to the Mars Rover's launch date (November 26, 2011) and its subsequent landing (August 6, 2012). 
}
\label{fig:curiosity}
\end{figure}

To study emerging entities, we analyze entities in a sample of online social media and news text streams spanning over 18 months. 
We focus on the entities' \emph{emergence patterns}, i.e., how an entity's exposure evolves between its first mention in online text streams, and the moment it is added to Wikipedia. 
We define an entity's \emph{emergence pattern} to be its ``document mention time series,'' i.e., the time series that represent the number of documents that mention the entity per day,\footnote{Because we are interested in broad and long-term patterns, our time series are at a granularity of days, not hours.} starting at its first mention in the stream, until it is incorporated into Wikipedia. 
An example time series is shown in Figure~\ref{fig:curiosity}, 
with the number of documents that mention \texttt{Curiosity} on the $y$-axis (the \emph{emergence volume}) 
and the time span between the entity's first mention in online text streams and the day it is added to Wikipedia on the $x$-axis (the \emph{emergence duration}).

The main findings of this paper are as follows. 
By clustering entity's emergence patterns, we find two kinds of regularity: entities that show a strong \emph{early burst} around the time of their introduction into public discourse, and \emph{late bursting} entities that exhibit a more gradual emergence. 
Furthermore, we find meaningful differences between how entities emerge in social media and news streams: entities that emerge in social media tend to transition more slowly from communicative \ac{CM} to cultural \ac{CM} than those that emerge in news streams. 
Finally, we show how different entity \emph{types} exhibit different emergence patterns; the fastest emerging entities are types that know shorter life-cycles such as devices (e.g., smartphones), and ``cultural artifacts'' (e.g., movies and music albums). 


\section*{Related Work}
\label{sec:relatedwork}

The concept ``collective memory'' was analyzed and advanced in the 1920s by \citet{halbwachs-cadres-1925,halbwachs1997memoire}. 
Since its introduction the concept has been studied in a variety of interdisciplinary fields, 
most notably in literature, history, and media~\cite[]{erll2008,9780195337426} 
but also in (experimental) psychology~\cite[]{doi:10.1080/09658210701811813,doi:10.1080/09658210701811912,doi:10.1080/09658210701806516}, e.g., by empirically studying the performance of remembering events of different members of a single social group~\cite[]{Brown1990OrganizationOP}. 

Wikipedia was first dubbed a global memory place where collective memories are built by \citet{Pentzold01052009}, with follow-up studies by \citet{keegan-breaking-2011} and \citet{ferron-collective-2011}. As \citet[p.~23]{ferron-collective-2012} puts it, ``Wikipedia's processes of discussion and article construction can be seen as the discursive formation of memory, or in other terms,
as the transition from \emph{communicative memory}, which is interactive, informal, nonspecialized, reciprocal, disorganized and unstable, to \emph{cultural memory}, which is formal, well organized and objective'' (our italics).

In the context of online collective memory, studies have revolved around automated methods for analyzing texts, e.g., studying temporal expressions in web documents has shown that we tend to remember the ``near past'' online~\cite[]{AuYeung:2011:SPR:2063576.2063755}. 
Wikipedia viewership statistics have provided insights into how current events trigger remembrance patterns of past events~\cite[]{Garcia-Gavilanese1602368}.
Other sources used for online collective memory studies include search engine query logs~\cite[]{campos2011temporal} and microblog services~\cite[]{Jatowt:2015:MTH:2736277.2741632}. 

Our work differs from previous work on collective memory in two important ways. We are the first to empirically study the transition from communicative \ac{CM} to collective \ac{CM} in terms of the entities that are mentioned in news and social text streams, before being included in Wikipedia. And we are the first to empirically study this transition at scale and across text streams and entity types, signifying an important difference from case studies that involve dramatic or traumatizing events, characteristic of the study of ``collective memory''~\cite[]{lipsitz-time-2001,neal-national-1998}.

\subsection*{Growth and development of Wikipedia}
Previous work on studying the expansion of Wikipedia through the addition of new pages usually studies the phenomenon from the perspective of Wikipedia itself, e.g., by analyzing how newly created articles fit in Wikipedia's semantic network, studying the relation between activity on talk pages and the addition of new content to articles, or by studying controversy and disagreement on new content through ``edit wars''~\cite[]{10.1371/journal.pone.0141892,Keegan01052013,10.1371/journal.pone.0038869,10.1371/journal.pone.0030091}. 

Emerging entities have emerged as object of study in the natural language processing and information retrieval communities. 
Different methods for identifying and linking unknown or emerging entities have been proposed~
\cite[]{NakasholeTW2013,Hoffart:2014:DEE:2566486.2568003,Lin:2012:NNP:2390948.2391045,voskarides-query-dependent-2014}.
\citet{graus-generating-2014} study the problem of predicting new concepts in social streams. \citet{Farber2016} study the specific challenges and aspects that come with linking emerging entities, while \citet{reinanda-document-2016} study the problem of identifying relevant documents for known and emerging entities as new information comes in, and \citet{graus-dynamic-2016} present a method for updating representations based on newly identified information.
Our work differs from the aforementioned studies in being observational in nature and its focus on temporal patterns. 

\section*{Research Questions} 
\label{sec:researchquestions}
In studying emergence patterns of entities, we apply different methods of grouping entities. 
First, we apply a burst-based unsupervised hierarchical clustering method to group entities by similarities in their emergence patterns. This allows us to answer the following question:

\begin{enumerate}[label=\textbf{RQ\arabic*},ref={RQ\arabic*},resume]
\item \acl{rq:clusters}\label{rq:clusters}
\end{enumerate}

\noindent
Next, we examine emerging entities in different types of text stream, viz.\ news and social media streams. 
In addition, we study the cross-pollination between the two types of streams, i.e., we study whether entities appear first in either of the streams, or whether they simultaneously appear in both. 
We answer the following question:

\begin{enumerate}[label=\textbf{RQ\arabic*},ref={RQ\arabic*},resume]
\item \acl{rq:substreams}\label{rq:substreams}
\end{enumerate}

\noindent
Finally, we characterize the emergence patterns of different types of entities. 
We leverage DBpedia, the structured counterpart of Wikipedia, to group emerging entities by their types, e.g., companies, athletes, and video games. 
We answer the following question:

\begin{enumerate}[label=\textbf{RQ\arabic*},ref={RQ\arabic*},resume]
\item \acl{rq:entities}\label{rq:entities}
\end{enumerate}

\section*{Data and Methods}

Our dataset spans 7.3 million time-stamped documents, with 36.2 million references to $n=79,482$ unique emerging entities, i.e., entities that did not have a Wikipedia entry at the time they were first mentioned in the corpus, but that did have one by the time the last document in the corpus was published. 

We create our custom dataset by extending the TREC-KBA StreamCorpus 2014\footnote{\url{http://trec-kba.org/kba-stream-corpus-2014.shtml}} with an additional set of annotations to Freebase entities (FAKBA1\footnote{\url{http://trec-kba.org/data/fakba1/}}). 
We then enrich the FAKBA1 dataset with links to Wikipedia, including for each link (i)~the creation date of the associated Wikipedia page, and (ii)~whether the Wikipedia page existed at the time the document was created. 
To encourage further research in emerging entities, we publicly release the tools needed to acquire the dataset used in this paper.\footnote{\url{https://github.com/graus/emerging-entities-timeseries}}

\subsubsection*{OOKBAT Dataset}
Our custom dataset is based on the TREC KBA StreamCorpus 2014, which comprises roughly 1.2 billion timestamped documents from global public news wires, blogs, forums, and shortened links shared on social media. 
It spans 572 days (October 7, 2011--May 1, 2013).

All (English) documents in the StreamCorpus have been automatically tagged for named entities with the Serif tagger~\cite[]{boschee2005automatic}, yielding roughly 580M tagged documents. 
\citet{fakba1} further automatically annotated these 580M documents with Freebase entities, resulting in the \emph{Freebase Annotations of TREC KBA 2014 Stream Corpus (FAKBA1) dataset}, which spans over 394M documents (Table~\ref{table1}, line~2). 
Because the Freebase used in FAKBA1 is dated after the StreamCorpus timespan, we can identify entities that appear in documents prior to being incorporated in Wikipedia. 

We take an entity's Wikipedia page creation date to be its time of transitioning from communicative to cultural \ac{CM}. 
To extract Wikipedia page creation dates for the Freebase entities present in FAKBA1, we leverage the available Wikipedia-mappings in Freebase. 
We then append the Wikipedia page creation dates (or \emph{entity timestamp}, denoted $e_{T}$) to each entity in the FAKBA1 dataset. 
In addition, we include the entity's ``age'' relative to the document timestamp ($doc_{T}$): the period in days between $e_{T}$ and $doc_{T}$, i.e., $e_{age} = e_{T} - doc_{T}$. 
The resulting dataset, FAKBA1, extended with the entity age and entity timestamp, is denoted \emph{Freebase Annotations of TREC KBA 2014 Stream Corpus with Timestamps (FAKBAT)} (Table~\ref{table1}, line~3). 

We retain only documents that contain entities with $e_{age} < 0$, i.e., emerging entities that are mentioned in documents dated before the entity's Wikipedia creation date. 
We denote the resulting subset of FAKBAT documents with emerging entities \emph{Out of Knowledge Base Annotations (with) Timestamps (OOKBAT)} (Table~\ref{table1}, line~4). 

To study an emerging entity's emergence patterns, we take two additional filtering steps.
First, we prune entities with creation dates more recent than the last document in our stream, to ensure the entities emerged in the timespan of our document stream. 
Next, we prune all entities that are mentioned in fewer than 5 documents. 
This yields our final dataset, which comprises 79,482 emerging entities (Table~\ref{table1}, line~5). 

\begin{table}[tb]
\centering
\caption{Descriptive statistics of our dataset acquisition. Coverage over preceding dataset in brackets. 
Looking at the second and third row in the table, we note that roughly two-thirds of the FAKBA1 entities can be mapped to Wikipedia. However, this portion represents 98\% of the mentions. 
The missing one-third were Freebase entities that had no links to Wikipedia, most notably, WordNet concepts and entities from the ``MusicBrainz'' knowledge base (i.e., artists, albums, and artists). 
The last two rows show that one in ten of the entities emerge during the span of the dataset, however, they constitute a mere 1\% of the mentions.}
\label{table1}
\small
\begin{tabular}{l@{ }l @{ } r@{ }r @{ } r@{ }r @{ } r@{ }r}
\toprule
& {\bf Dataset} & \multicolumn{2}{c}{\bf \# entities} & \multicolumn{2}{c}{\bf \# mentions} & \multicolumn{2}{c}{\bf \# documents} \\ 
\midrule
1. & \emph{TREC KBA} & N/A & & N/A & & 579,838,246 &  \\ 
2. & \emph{FAKBA1} & 3,272,980 && 9,423,901,114 && 394,051,027 & (68.0\%) \\ 
\midrule
3. & FAKBAT & 2,254,177 & (68.9\%) & 9,221,204,641 & (97.8\%) & 394,051,027 & (100\%) \\ 
4. & OOKBAT & 225,291 & (10.0\%) & 94,929,292 & (1.0\%) & 23,896,922 & (6.1\%) \\
\midrule
5. & Emerging entities & 79,482 & (35.3\%) & 36,242,096 & (38.2\%) & 7,291,700 & (30.5\%) \\ 
\bottomrule
\end{tabular}
\end{table}

\subsubsection*{Entity types}
To study entity types for RQ3, we map emerging entities to their respective classes assigned in the DBpedia ontology,\footnote{\url{http://mappings.dbpedia.org/server/ontology/classes/}} e.g., the entity \texttt{Barack\_Obama} is mapped to the Person, Politician, Author, Award Winner classes.
Out of the 79,482 emerging entities in our dataset, we have 39,713 class-mappings (a coverage of 50.0\%). 

\subsubsection*{Entity popularity}
As a proxy for an entity's popularity, we extract Wikipedia pageview statistics. 
We extract the total number of pageviews each entity received during 2015. 
We choose to use the pageview counts of a year that falls outside of the timespan of our dataset so as to minimize the effects of timeliness. 

\subsection*{Time series clustering} 
The core unit in our analysis are so-called \emph{emergence patterns}, i.e., time series that represent the number of documents that mention an entity over time. 
To answer our first research question, \ac{rq:clusters}, we apply clustering to group entities with similar emergence patterns. 
Clustering time series consists of three steps: 
First, we normalize the time series, as they might span very different periods of time. 
Next, we measure the similarities between time series. 
And third, we apply hierarchical agglomerative clustering. 

\subsubsection*{Normalization} 
\label{subsec:norm}
The entities' time series we study here are characterized by several properties.
First, they are of variable length: 
some entities may take days to transition, others take months. 
Second, the time series in our dataset are temporally unaligned: 
each time series starts at the timestamp of the article that first mentions the entity, and ends at the entity's Wikipedia page creation date. 
To be able to visualize the time series of variable lengths, we linearly interpolate the time series to have equal length~\cite[]{ijca2012clustering}. 
Furthermore, since we are not interested in the absolute differences in document volumes or entity popularity when visualizing the time series clusters, we standardize our time series by subtracting the mean and dividing by the standard deviation~\cite[]{Vlachos:2004:ISP:1007568.1007586}. 

\subsubsection*{Similarity} 
\begin{figure}[t]
\centering
\includegraphics[width=\linewidth]{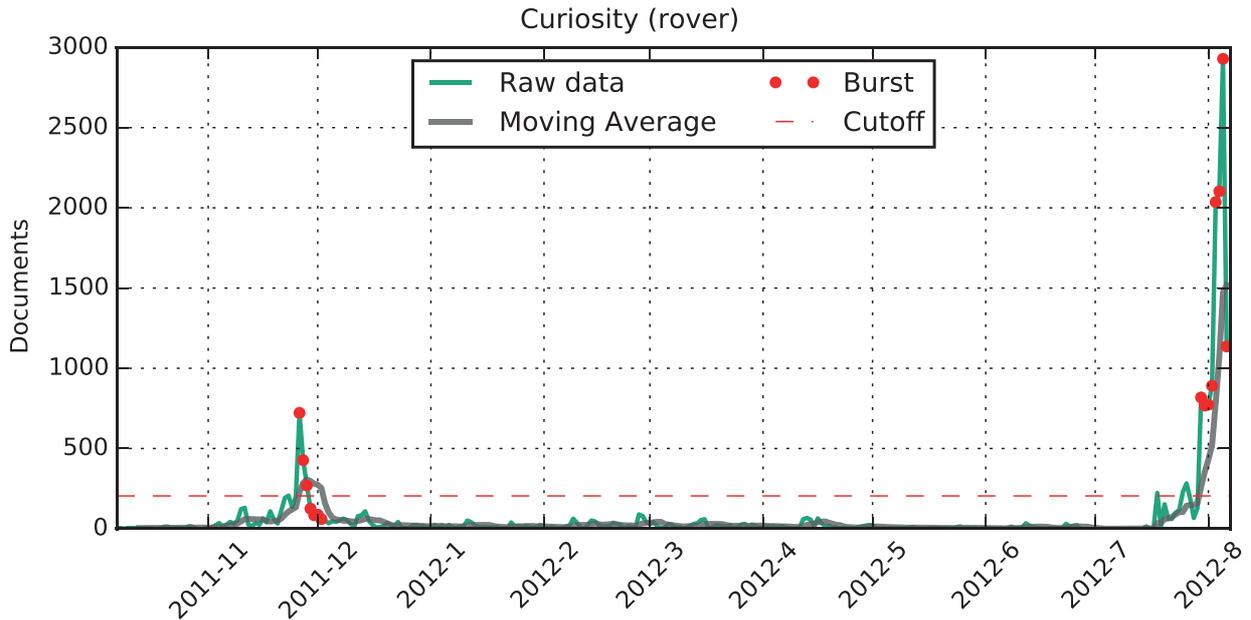}
\caption{Detected bursts of \texttt{Curiosity (rover)}'s time series. The bursts correspond to the earlier described launch and landing of the Mars Rover (see also Figure~\ref{fig:curiosity}).}
\label{fig:curiositybursts}
\end{figure}

Typically, time series similarity metrics rely on fixed-length time series, and may leverage seasonal or repetitive patterns~\cite[]{WarrenLiao20051857}. 
Since our time series are of variable length, and not temporally aligned, common time series similarity metrics such as Dynamic Timewarping (DTW) are not suitable~\cite[]{Bemdt94usingdynamic}. 
Furthermore, we are interested in periods in which the exposure of an entity in public discourse increases or changes. 
These ``bursts'' may be correlated to real-world activity and events around the entity. 
To address the nature of the time series, and our focus on bursts we employ BSim~\cite[]{Vlachos:2004:ISP:1007568.1007586} (Burst Similarity) as our similarity metric. 
BSim relies on measuring the overlap between bursts of different time series. 
To detect these bursts we compute a moving average of each emerging entity time series ($T_{e}$), denoted $T_{e}^{MA}$. 
We set parameter $w$ (the size of the rolling window) to 7 days. 
Bursts are the points in $T_{e}^{MA}$ that surpass a cutoff value ($c$). 
We set $c=1.5 \cdot \sigma_{MA}$, where $\sigma_{MA}$ is the standard deviation of $MA$. 
The parameter choices for $w$ and $c$ are in line with previous work~\cite[]{Vlachos:2004:ISP:1007568.1007586}. 
Figure~\ref{fig:curiositybursts} shows an example time series ($T_{e}$), with the bursts detected for the previously shown \texttt{Curiosity (rover)}.
The detected bursts correspond to the earlier mentioned launch and landing of the Mars rover.

\subsubsection*{Hierarchical agglomerative clustering} 
To cluster time series, we compute pair-wise similarities between all time series, and yield Similarity Matrix $SM$. 
We then apply $L_2$ normalization to $SM$, and convert it to a distance matrix $DM$ ($DM = 1 - SM$). 
Finally, we apply hierarchical agglomerative clustering (HAC) on $DM$ using the \texttt{fastcluster} package~\cite[]{JSSv053i09}. 
As our linkage criterion, we employ \emph{Ward's method}~\cite[]{doi:10.1080/01621459.1963.10500845}.

\subsection*{Analysis}
\label{subsec:analysis}

Given our grouping methods (clustering, by entity type, by stream type), we apply two methods to analyze the resulting groups of emerging entities: 
(i)~visualization of group signatures, and 
(ii)~descriptive statistics that reflect properties of the underlying time series. 

\subsubsection*{Visualization}
To compare groups of emerging entities, we visualize their so-called \emph{group signatures}, i.e., the average of all time series that belong to a group. 
See Figure~\ref{fig:global} for an example group signature of all emerging entities in our dataset ($n$ = 79,482). 
As described above, the time series (may) differ in length, and are not temporally aligned. 
To visualize the time series, we linearly interpolate each to the (overall) longest emergence duration, effectively ``stretching'' them to have equal length. 
Next, we align them in relative duration, i.e., we overlay each entity's first and last mention at the start and end of the $x$-axis, respectively. 

\subsubsection*{Descriptive statistics}
Visualizing group signatures does not paint the full picture.  
More fine-grained aspects of emergence, e.g., the average \emph{emergence duration} (the time between an entity's first mention in the text stream, and its subsequent incorporation into Wikipedia), or \emph{emergence volume} (the number of documents that mention the entity before it is incorporated into Wikipedia) disappear through our visualization method. 
To study these aspects, we describe the time series groups using different features that reflect the emergence behavior of the group. 
For an overview of the descriptive statistics that we consider, see Table~\ref{tab:tsfeatures}.

\begin{table}[t]
\centering
\caption{Descriptive statistics used for analyzing and comparing different groupings of emerging entities. We distinguish between \emph{time series statistics} (top) and \emph{burst statistics} (bottom). 
All statistics are computed for the period ranging from the emerging entity's first mention in the corpus to the creation date of the Wikipedia page devoted to it.}
\label{tab:tsfeatures}
\begin{tabular}{ll}
\toprule
\textbf{Emergence volume} & Number of documents that mention the entity \\
\textbf{Emergence duration} & Number of days from first mention to incorporation \\
\textbf{Emergence velocity} & $\frac{Volume}{Duration}$ (average number of documents per day) \\
\midrule
\textbf{Bursts number} & Total number of bursts \\
\textbf{Bursts duration} & Normalized average durations of bursts (i.e., bursts widths) \\
\textbf{Bursts value} & Normalized average heights of burst (i.e., bursts heights) \\
\bottomrule
\end{tabular}
\end{table}

\section*{Results}
\label{sec:results}

In this section we present the analyses that answer our research questions. 


\subsection*{RQ1: Emergence patterns}
\label{sec:rq1}

\begin{figure}[t]
\centering
\includegraphics[width=\linewidth]{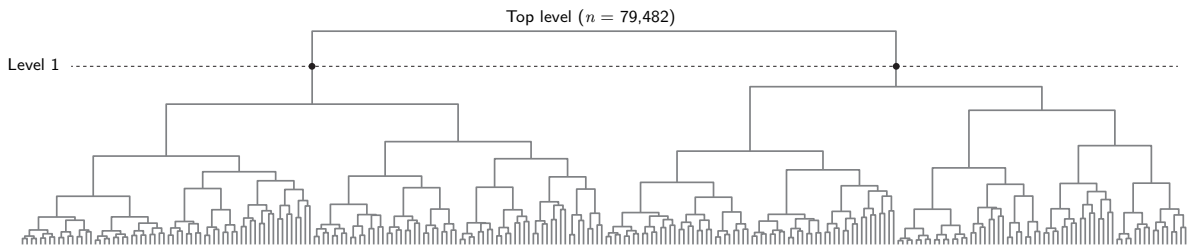}
\caption{Dendrogram resulting from applying hierarchical agglomerative clustering using BSim similarity~\cite[]{Vlachos:2004:ISP:1007568.1007586}, on our corpus of emerging entity time series ($n$ = 79,482). 
The cutoff-points at which we analyze the clusters are denoted Top level, and Level 1 (2 clusters). 
For clarity, the tree is truncated by showing no more than 7 levels of the hierarchy. }
\label{fig:dendrogram}
\end{figure}

Figure~\ref{fig:dendrogram} shows a cluster tree that results from clustering the time series distance matrix. 
At its highest level, the tree shows two distinct clusters, each of which is broken down into multiple smaller sub-clusters.  
In the following section, we study the \emph{global emergence patterns}, by taking all time series at the root of the tree (\emph{Top level} in Figure~\ref{fig:dendrogram}), and next, the two main clusters (\emph{Level 1} in Figure~\ref{fig:dendrogram}).

\subsubsection*{Global emergence pattern}
\begin{figure}[t]
\centering
\includegraphics[trim={7.5cm 0 0 0},clip,height=1.5in]{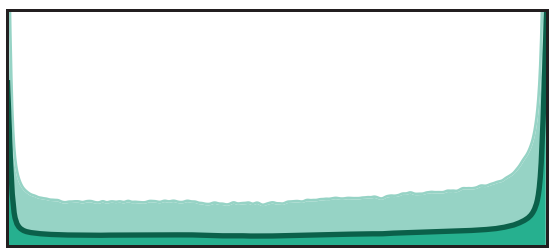}
\caption{Global cluster signature (of all emerging entities) where $n$ = 79,482. 
That is, the top level in the dendrogram in Figure~\ref{fig:dendrogram}. 
The axes are not labeled since all time series values (i.e., document counts) are standardized, and the series are linearly interpolated to have equal length. 
The solid line represents the cluster signature (i.e., the average time series), the lighter band represents standard deviation.}
\label{fig:global}
\end{figure}

Figure~\ref{fig:global} shows how both the emerging entities' introduction into public discourse (the first mention at the left-most side of the plot) and subsequent incorporation into cultural \ac{CM} (the right-most side of the plot) occur in bursts of documents, i.e., overall, the largest number of documents that mention a newly emerging entity are either at the start or at the end of their time series. 
This can be explained as follows. The entrance into public discourse represents the first emergence of an entity, whereas being added to cultural \ac{CM} is likewise likely to happen in a period of increased attention, e.g., a real world event that puts the entity in public discourse.  Between these two bursts, the number of documents that mention the entity seems to increase gradually as time progresses, suggesting that on average, the number of documents that mention a new entity, and thus the attention the entity receives in public discourse increases over time before it reaches ``critical mass.''

Turning to the descriptive statistics in Table~\ref{tab:global}, it takes 245~days on average for an entity to emerge, but with large variations between entities, motivating our clustering approach.
On average, an entity is associated with multiple bursts (3.8), indicating that entities are likely to resurface multiple times in public discourse before being deemed important enough to transition into cultural \ac{CM}. 

\begin{table}[t]
\centering
\small
\caption{Global time series and burst descriptive statistics.}
\label{tab:global}
\begin{tabular}{r@{ $\pm$ }r r r@{ $\pm$ }r r r@{ $\pm$ }r r}
\toprule
\multicolumn{3}{c}{duration (\# days)} & \multicolumn{3}{c}{volume (\# docs)} & \multicolumn{3}{c}{velocity (docs/day)} \\
mean & std & med. & mean & std & med. & mean & std & med. \\
\midrule
245 & 153 & 221 & 183 & 1,180 & 32 & 0.87 & 5.6 & 0.19 \\
\midrule
\multicolumn{3}{c}{n\_bursts} & \multicolumn{3}{c}{bursts durations} & \multicolumn{3}{c}{bursts values} \\
\midrule
3.8 & 2.62 & 3 & 0.03 & 0.03 & 0.02 & 0.03 & 0.08 & 0.02 \\
\bottomrule
\end{tabular}
\end{table}

\subsubsection*{Clusters at level~1 in Figure~\ref{fig:dendrogram}: Early vs.\ late bursts}
\label{subsec:mainclusters}

\begin{figure}[t]
\centering
\includegraphics[trim={0 0 0 0},clip,width=\linewidth]{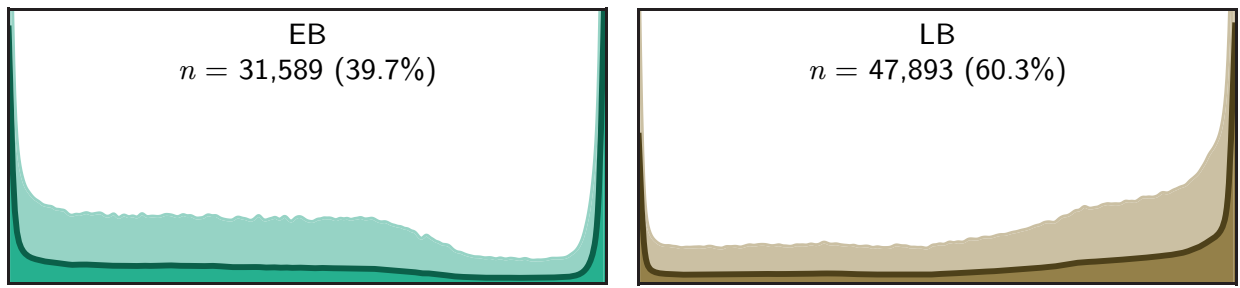}
\caption{Cluster signatures of the \acl{EB} entities (left plot) and \acl{LB} entities (right plot) clusters, denoted Level 1 in Figure~\ref{fig:dendrogram}. 
} 
\label{fig:clusters}
\end{figure}

In our first attempt at uncovering distinct patterns in which collective remembrance happens, we study the two main clusters at Level~1 of the cluster tree (Figure~\ref{fig:dendrogram}). 
The resulting cluster signatures are shown in Figure~\ref{fig:clusters}. 

Much like the global cluster signatures in the previous section, the Level~1 clusters show two main bursts: the \emph{initial burst} around the first mention, and the \emph{final burst} around the time an entity is added to Wikipedia. 
Howeve, the left cluster, which we call \emph{\acf{EB} entities}, is characterized by a stronger initial burst, with the majority of the documents that mention the entity concentrated at the time when the entity surfaces in communicative \ac{CM}. 
This suggests that the cluster contains new entities that suddenly emerge and experience a (brief) period of lessened attention, before transitioning into the collective's \ac{CM}. 
The right cluster, which we denote as \emph{\acf{LB} entities}, shows a more gradual pattern in activity towards the point at which the entity is incorporated into cultural CM, much like we saw in the global signature. 

We note two main differences between the group signatures of the \ac{EB} and \ac{LB} entities in Figure~\ref{fig:clusters}. 
First, the distribution of documents between the initial and final burst. 
The \ac{EB} entities show a more ``abrupt'' final burst: 
the majority of the documents are in the wake of the initial burst, i.e., at the left-hand side of the plot, then, the document volume gradually winds down, before it finally seems to abruptly transition into the final burst at the right hand-side of the plot. 
In contrast, the \ac{LB} entities cluster shows a relatively subtle initial burst, which likewise quiets down, followed by a gradual increase of document volume that leads up to the final burst. 

A second difference is the height difference between the initial and final bursts. 
The \ac{EB} cluster shows roughly equally high initial and final bursts; the \ac{LB} cluster shows a substantially smaller initial burst, which suggests the introduction into public discourse is more subtle than its addition to Wikipedia. 

We turn to the clusters' descriptive statistics in Table~\ref{tab:clusterstats}.
We first test for statistical significance in the differences between the cluster statistics.
We perform a Kruskal-Wallis one-way analysis of variance test, and follow this omnibus test with a post-hoc test using Dunn's multiple comparison test (with $p$-values corrected for family-wise errors using Holm-Bonferroni correction). We find that all differences are statistically significant at the $\alpha$ = 0.05 level. 

Table~\ref{tab:clusterstats} shows 
\ac{LB} entities emerge more slowly (259 days) than \ac{EB} entities (224 days). 
\ac{LB} entities also receive more exposure during emergence (225 vs.\ 118 documents for \ac{EB} entities). 
The shorter emergence duration and lower volumes seen with the \ac{EB} entities suggest they represent more popular, timely, or ``urgent'' entities, that will be incorporated quickly after emerging in public discourse, e.g., large-scale events and popular entities. 
The descriptive statistics of the \ac{LB} entities on the other hand suggests less timely or urgent entities. 
The burst statistics confirm this view of slower, less timely \ac{LB} entities, and more urgent, faster \ac{EB} entities: 
the average burst heights of \ac{EB} entities are higher, suggesting \ac{LB} entities see a more evenly spread volume of documents that mention them. 
Furthermore, \ac{EB} entities show fewer bursts (3.22 vs.\ 4.12, on average). 

And indeed, the EB entities that occur most frequently in our dataset include many ``central'' entities related to popular culture, e.g., products such as \texttt{Xbox One} (121,813 mentions), movies, e.g., \texttt{The Twilight Saga: Breaking Dawn - Part 2} (124,222 mentions), and news events, e.g., \texttt{Disappearance of Lisa Irwin} (15,917 mentions). 
The most frequent LB entities on the other hand include more obscure, long-tail, or niche entities: most notably people, e.g., \texttt{Jeffrey Chiesa} (31,560 mentions), \texttt{Sergio Ermotti} (22,274 mentions), and \texttt{James Rolfe (filmmaker)} (15,797 mentions).

\begin{table}[t]
\small
\centering
\caption{Comparison of \acl{EB} and \acl{LB} entities clusters statistics. }
\label{tab:clusterstats}
\begin{tabular}{l @{} r r@{ $\pm$ }r r r@{ $\pm$ }r r r@{ $\pm$ }r r}
\hline
& proportion & \multicolumn{3}{c}{duration (\#days)} & \multicolumn{3}{c}{volume (\#docs)} & \multicolumn{3}{c}{velocity (docs/day)} \\
& & mean & std & med. & mean & std & med. & mean & std & med. \\
\hline
\ac{EB} & 0.40 & 224 & 146 & 195 & 118 & 804 & 22 & 0.70 & 6.45 & 0.15 \\
\ac{LB} & 0.60 & 259 & 156 & 238 & 225 & 1,371 & 42 & 0.99 & 4.96 & 0.23 \\
\hline
& & \multicolumn{3}{c}{n\_bursts} & \multicolumn{3}{c}{burst durations} & \multicolumn{3}{c}{burst values} \\
\hline
\ac{EB} & 0.40 & 3.32 & 2.20 & 3  & 0.03 & 0.03 & 0.02 & 0.05 & 0.11 & 0.02 \\
\ac{LB} & 0.60 & 4.12 & 2.82 & 4 & 0.03 & 0.03 & 0.02 & 0.03 & 0.05 & 0.01 \\
\hline
\end{tabular}
\end{table}

\subsubsection*{Summary}
In this section, we have answered our first research question: ``\acl{rq:clusters}'' We performed hierarchical clustering using a burst similarity-metric of the emerging entity time series and discovered two distinct emergence patterns: 
\acl{EB} entities and \acl{LB} entities. 
Our visual inspection of the cluster signatures suggest \ac{LB} entities emerge more slowly, i.e., build up attention more slowly before transitioning from communicative into cultural \ac{CM}, whereas \ac{EB} entities are associated with more sudden and higher bursts of activity, prior to transitioning into cultural \ac{CM}.
We find that the two clusters differ substantially and significantly in their cluster signature and descriptive statistics.


\subsection*{RQ2: Emerging entities in social media and news}
\label{sec:rq2}

In this section we answer our second research question: ``\acl{rq:substreams}''
In the previous section we have shown that 79,482 entities emerge in the combined news and social media streams. 
By splitting out these entities by stream, we find 51,095 of these entities emerge in the news stream (i.e., are mentioned in the news stream), similar to the number of entities that emerge in the social media stream, at 51,356. 
Finally, 30,148 of the emerging entities are mentioned in both streams before being incorporated into Wikipedia.

\subsubsection*{Global: News vs.\ social}
First, we compare the emergence patterns of entities in news and social streams. 
We apply the same hierarchical clustering method from the previous section on the two subsets of entities that emerge in news and social media streams (where $n_{news}$ = 51,095 and $n_{social}$ = 51,356). 

\begin{figure}[t]
\centering
\includegraphics[width=.85\linewidth]{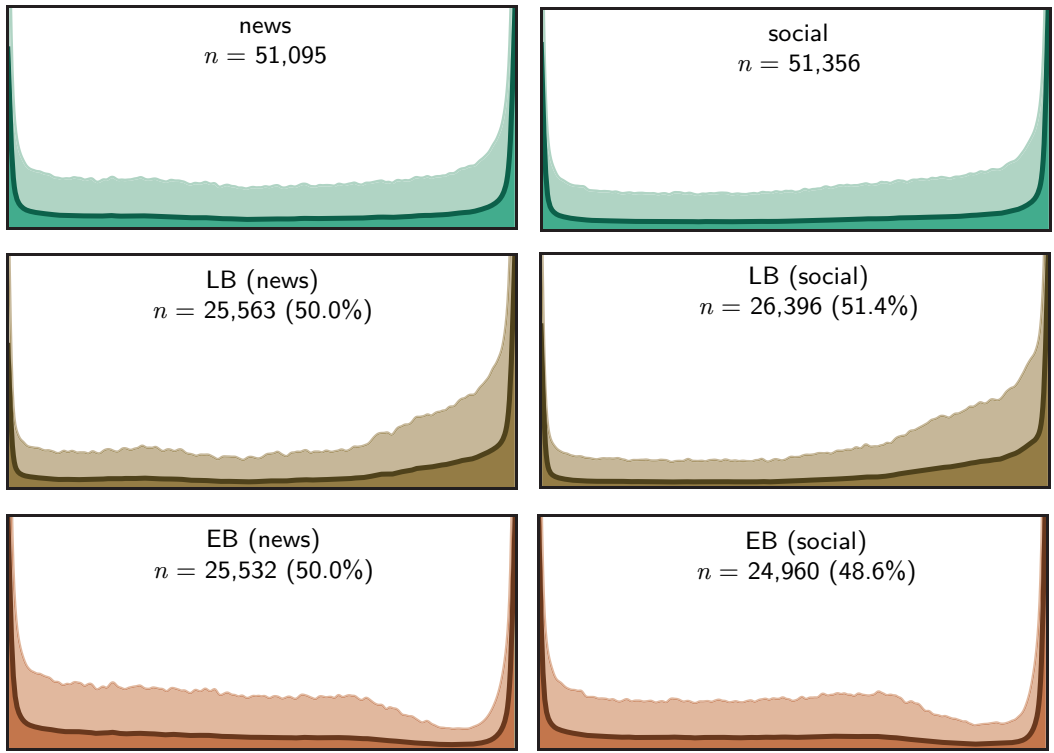} 
\caption{News vs.\ Social stream cluster signatures. 
The top row shows the global cluster signature of the news (left) and social (right) streams. 
The bottom two rows show the signatures of the \acl{LB} and \acl{EB} entity clusters for each stream (news left, social right).}
\label{fig:substreamclustercomparison}
\end{figure}

Figure~\ref{fig:substreamclustercomparison} first shows the global emergence patterns (top row, in green), which are largely the same in the two streams and highly similar to the global patterns studied in the previous section. 
The bottom two rows of Figure~\ref{fig:substreamclustercomparison} show that both streams exhibit groups that are similar to the \acl{EB} and \acl{LB} entities shown in Figure~\ref{fig:clusters}. 
Looking at the top row, however, shows that entities that emerge in news have slightly more of their emergence volume mass after the initial burst, compared to the corresponding pattern of the social media stream, which exhibits a more gradual increase in emergence volume towards the final burst at the right-hand side of the plot. 
This may be attributed to the slightly higher proportion of \acl{EB} entities in the news stream, which has 50.0\% of its entities falling in this cluster, while the social media stream has 48.6\%. 

\subsubsection*{Who's first?}
Of the 79,482 entities that emerge in the 18 month period our dataset spans, 45,678 appear in both the news and social media stream before they transition to cultural \ac{CM};  
9,681 entities are mentioned exclusively in the news stream, and never appear in social media (\texttt{news-only}) before transitioning into cultural CM. 
Finally, 23,096 appear only in the social media stream (\texttt{social-only}). 
In Table~\ref{tab:emergewhere}, we compare entities as they emerge in different streams.

\begin{table}[t]
\centering
\caption{Emergence features for our five groups of entities: 
entities that emerge in both streams, but first in the news stream (\texttt{news-first}), 
entities that emerge in both streams, but first in the social media stream (\texttt{social-first}), 
entities that emerge in both streams on the same day (\texttt{same-time}), 
entities that emerge only in the news stream (\texttt{news-only}), and finally, 
entities that emerge only in the social media stream (\texttt{social-only}).
}
\label{tab:emergewhere}
\small
\begin{tabular}{l r@{ $\pm$ }r r r@{ $\pm$ }r r r@{ $\pm$ }r r}
\toprule
stream & \multicolumn{3}{c}{duration (\#days)} & \multicolumn{3}{c}{volume (\#docs)} & \multicolumn{3}{c}{velocity (docs/day)} \\
& mean & std & med. & mean & std & med. & mean & std & med. \\
\midrule
news first   &  298 & 139 & 305 & 123 & 291 & 53 & 0.58 & 1.59 & 0.21 \\
social first &  281 & 157 & 276 & 182 & 445 & 74 & 0.95 & 3.22 & 0.32 \\
same time    &  197 & 147 & 163 & 192 & 662 & 67 & 2.87 & 23.59 & 0.51 \\
only news    &  250 & 152 & 216 & 415 & 2,215 & 65 & 1.45 & 6.45 & 0.35 \\
only social  &  214 & 148 & 190 & 33 & 134 & 12 & 0.41 & 2.60 & 0.08 \\
\bottomrule
\end{tabular}
\end{table}

Of the 45,678 entities that emerge in both streams, the majority appears in the social media stream before they appear in the news stream. 
This may be explained by the different nature of the publishing cycles of the two streams; whereas news stories need to be checked and edited before being published, social media follows a more unedited and direct publishing cycle. 

The entities that appear in social media first (\texttt{social-first}), cover 64.9\% ($n$ = 29,665) of the entities that emerge in both streams. 
Interestingly, entities that emerge in news first, subsequently appear in social media streams slower than vice versa: on average 66 days for the former, and 49 days for the latter. 
A relatively small number of entities is mentioned in both streams on the same day (\texttt{same\-time}): 8.7\% ($n$ = 3,967). 
Such entities are expected to being more urgent and central, as they appear more widely in public discourse. 
This group's shortest emergence durations and highest velocities, support this view of entities that play a more central role in public discourse. 
And indeed, looking at the entities that appear in this set, we see a large number of news events-related entities, e.g., \texttt{12-12-12: The Concert for Sandy Relief}, \texttt{2013 Alabama bunker hostage crisis}, and \texttt{Suicide of Jacintha Saldanha}.

\subsubsection*{Summary}
News and social media streams show broadly similar emergence patterns for entities but the population and the behavior of entities emerging in news and social differ significantly. 
Entities are slower on average in emerging in social media streams, and entities that appear in both streams on the same day are the fastest to transition to cultural \ac{CM}.


\subsection*{RQ3: Emergence patterns of different entity types}
\label{sec:rq3}

In this section we answer our third research question: ``\acl{rq:entities}''
We compare the descriptive statistics of different entity types in our dataset, to assert whether different types exhibit different emergence patterns.

\subsubsection*{Entity types: temporal patterns}
\begin{figure}[t]
\centering
\includegraphics[width=.95\linewidth]{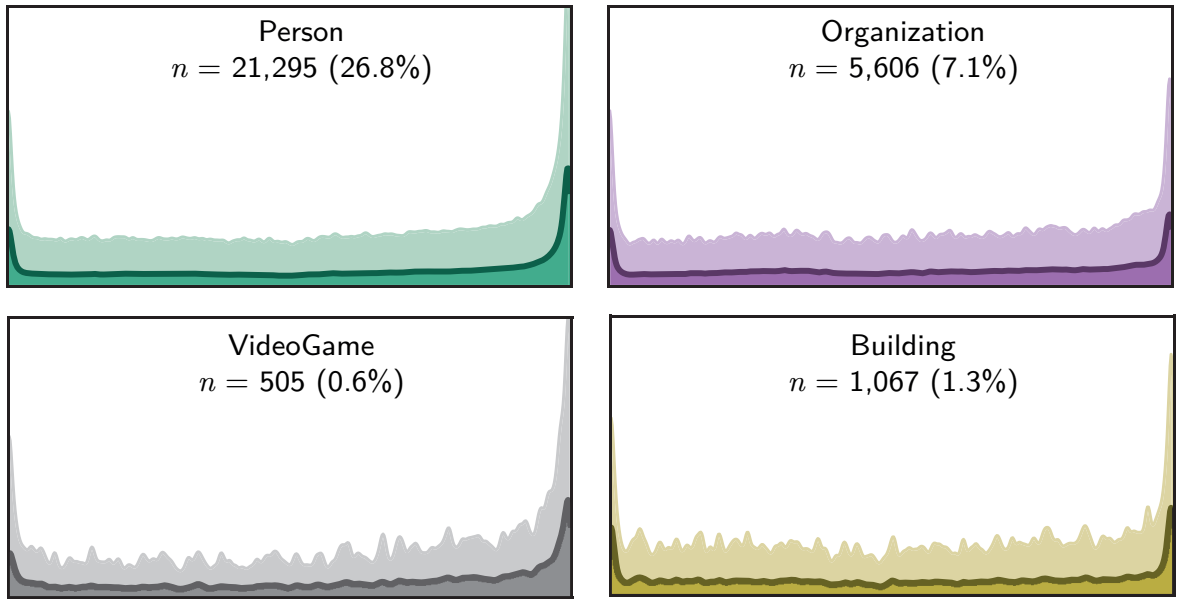} 
\caption{Type signatures of the Person, Organization, VideoGame and Building types. Even though the number of entities per type differ substantially, the signatures show similar patterns.}
\label{fig:classplots}
\end{figure}

First, we study the descriptive statistics per entity type. Table~\ref{tab:entitytypes} provides an overview of the most frequent entity types in our dataset (i.e., all entity types with a frequency of $\geq 400$). 
We find that the entity type signatures are very similar to the global pattern of Figure~\ref{fig:global}, which suggests the time series are highly variable within an entity type. 
See Figure~\ref{fig:classplots} for an example of two common entity types (top row) and two less frequently emerging types (bottom row): whereas the signature becomes smoother as the number of mentions increases, the overall pattern is highly similar across the four types. 

\begin{table}[t]
\small
\centering
\caption{Descriptive statistics per entity type (for types that occur $\geq 400$ times in our dataset). }
\label{tab:entitytypes}
\begin{tabular}{l@{}r @{~} r@{ $\pm$ }r @{~} r @{~} r@{ $\pm$ }r @{~} r @{~} r@{ $\pm$ }r @{~} r}
\toprule
stream & n\_samples & \multicolumn{3}{c}{duration (\#days)} & \multicolumn{3}{c}{volume (\#docs)} & \multicolumn{3}{c}{velocity (docs/day)} \\
& & mean & std & med.\ & mean & std & med.\ & mean & std & med.\ \\
\midrule
Person & 21,295 & 270 & 151 & 254 & 243 & 692 & 71 & 1.03 & 3.32 & 0.32 \\
Athlete & 8,018 & 260 & 150 & 235 & 264 & 674 & 76 & 1.05 & 2.45 & 0.37 \\
InformationEntity & 7,847 & 242 & 154 & 210 & 294 & 1,923 & 90 & 1.42 & 6.53 & 0.51 \\
CreativeWork & 7,795 & 243 & 154 & 211 & 294 & 1,928 & 90 & 1.42 & 6.54 & 0.52 \\
Organization & 5,606 & 279 & 153 & 270 & 335 & 1,812 & 71 & 1.40 & 14.44 & 0.31 \\
Place & 3,689 & 274 & 149 & 273 & 122 & 448 & 33 & 0.48 & 1.61 & 0.16 \\
Company & 2,536 & 284 & 156 & 275 & 462 & 1,964 & 108 & 1.98 & 20.88 & 0.47 \\
MusicalWork & 2,474 & 218 & 148 & 181 & 170 & 533 & 78 & 1.13 & 2.23 & 0.49 \\
Movie & 2,033 & 267 & 154 & 247 & 279 & 1,322 & 87 & 1.20 & 6.57 & 0.42 \\
OfficeHolder & 1,929 & 287 & 158 & 284 & 210 & 476 & 73 & 0.85 & 1.88 & 0.31 \\
MusicGroup & 1,649 & 293 & 150 & 289 & 221 & 393 & 86 & 0.95 & 1.97 & 0.34 \\
Artist & 1,624 & 299 & 152 & 302 & 240 & 564 & 75 & 0.95 & 2.19 & 0.30 \\
ArchitecturalStructure & 1,591 & 279 & 149 & 284 & 133 & 436 & 36 & 0.47 & 1.14 & 0.18 \\
PopulatedPlace & 1,521 & 262 & 145 & 244 & 119 & 481 & 31 & 0.53 & 2.15 & 0.15 \\
Building & 1,067 & 281 & 150 & 290 & 125 & 374 & 34 & 0.43 & 0.98 & 0.17 \\
TelevisionShow & 1,043 & 229 & 158 & 193 & 228 & 503 & 87 & 1.33 & 3.05 & 0.57 \\
WrittenWork & 959 & 267 & 147 & 245 & 307 & 864 & 88 & 1.26 & 2.99 & 0.44 \\
EducationalInstitution & 915 & 290 & 144 & 308 & 108 & 564 & 30 & 0.41 & 2.21 & 0.12 \\
Software & 769 & 250 & 156 & 221 & 732 & 5,413 & 192 & 3.04 & 16.33 & 0.92 \\
School & 554 & 280 & 142 & 305 & 56 & 158 & 25 & 0.29 & 2.24 & 0.11 \\
Book & 524 & 272 & 147 & 263 & 286 & 827 & 93 & 1.17 & 3.03 & 0.44 \\
VideoGame & 505 & 229 & 150 & 198 & 381 & 657 & 189 & 2.08 & 3.24 & 1.03 \\
DesignedArtifact & 409 & 217 & 149 & 187 & 1,420 & 7,142 & 214 & 7.04 & 20.42 & 1.39 \\
Infrastructure & 403 & 271 & 146 & 269 & 127 & 560 & 34 & 0.47 & 1.42 & 0.17 \\
\texttt{null} & 39,807 & 225 & 151 & 196 & 98 & 861 & 15 & 0.58 & 3.70 & 0.10 \\
\bottomrule
\end{tabular}
\end{table}

Turning to Table~\ref{tab:entitytypes}, 
we note the \texttt{null} class, i.e., entities that are not assigned an entity type in DBpedia exhibit very low emergence volumes (98 documents on average). This may be explained by their nature: long-tail, or unpopular entities are more likely to not have a class assigned in the DBpedia ontology. 

Second, we note a group of ``fast'' emerging entity types with short emergence durations and/or high emergence velocities, e.g., \texttt{DesignedArtifact}, \texttt{CreativeWork}, \texttt{Musi\-cal\-Work}, and \texttt{VideoGame}, consider, e.g., the \texttt{DesignedArtifacts} emergence velocity, at 217 days with over 7 documents a day on average. 
This type includes entities such as devices and products, e.g., smartphones, tablets, and laptops. 
The relatively fast transitioning speed may be explained by their nature: they have short ``life-cycles'' and may be superseded or replaced at high frequencies. 
Consider, e.g., the release or announcement of a new smartphone: this event typically generates a lot of attention in a short timeframe, which may result in a fast transition into cultural \ac{CM}.  
Similar to the \texttt{DesignedArtifact}-type, \texttt{CreativeWork}s (including, e.g., \texttt{MusicalWork}, \texttt{WrittenWork}, \texttt{Movie}) share this characteristic: they play a central but short-lived role in public discourse. 

Third, the ``slower'' entities, i.e., those with longer emergence durations and lower emergence volumes, are largely person types such as \texttt{Writer}, \texttt{Artist}, and political figures (\texttt{Off\-ice\-Hol\-der}), but also \texttt{School} and \texttt{EducationalInstitution}, and geographical entities (e.g., \texttt{Building}, \texttt{ArchitecturalStructure}, \texttt{Place}, and \texttt{Popula\-ted\-Place}). 
These entities may have longer life-cycles and a more gradual ``rise to fame'' by their nature, and have a less central role in public discourse. 
Consider, e.g., politicians who generally have a long and gradual career and are more likely to first emerge in local media. Similarly, an opening of a new school building may emerge in regional news, but is unlikely to be globally and widely reported. 

To better understand the difference between ``fast'' and ``slow'' entities, we examine the popularity of entities. Table~\ref{tab:typepopularity} lists the average number of pageviews received per entity in 2015, per type. 
Looking at the ranking, we note how ``faster'' emerging entity types remain more popular over time: 
types that are associated with short emergence durations and high velocities all fall in the top 10 (ranks 3, 4, and 9, for \texttt{VideoGame}, \texttt{CreativeWork}, and \texttt{DesignedArtifact}, respectively), whereas slower types reside in lower ranks in the table, e.g., rank 19, 22, and 24 for \texttt{Building}, \texttt{EducationalInstitution} and \texttt{School}, respectively. 

\begin{table}[t]
\centering
\small
\caption{``Popularity'' (number of pageviews) of each entity in our dataset, aggregated per entity type. Ranked in descending order. }
\label{tab:typepopularity}
\begin{tabular}{rlr@{ $\pm$ }rr}
\toprule
Rank & Type & \multicolumn{2}{c}{Mean $\pm$ std} & Median \\
\midrule
1 & Movie & 98,387 & 322,166 & 14,352 \\
2 & TelevisionShow & 97,098 & 309,172 & 9,765 \\
3 & VideoGame & 51,236 & 166,802 & 11,852 \\
4 & CreativeWork & 50,024 & 213,490 & 5,716 \\
5 & InformationEntity & 49,704 & 212,816 & 5,634 \\
6 & Software & 43,657 & 149,499 & 9,582 \\
7 & MusicGroup & 38,883 & 133,336 & 5,400 \\
8 & Artist & 35,607 & 122,032 & 4,116 \\
9 & DesignedArtifact & 29,830 & 82,081 & 7,191 \\
10 & Book & 18,248 & 109,400 & 3,126 \\
11 & WrittenWork & 14,227 & 86,637 & 1,801 \\
12 & Person & 13,772 & 77,791 & 1,568 \\
13 & MusicalWork & 10,443 & 25,009 & 3,523 \\
14 & Athlete & 9,415 & 41,887 & 1,545 \\
15 & Organization & 9,003 & 45,140 & 1,816 \\
16 & Company & 7,624 & 21,371 & 2,566 \\
17 & OfficeHolder & 3,763 & 16,167 & 958 \\
18 & ArchitecturalStructure & 3,189 & 16,978 & 1,042 \\
19 & Building & 3,180 & 20,106 & 987 \\
20 & Infrastructure & 2,813 & 6,769 & 1,085 \\
21 & Place & 2,339 & 12,649 & 827 \\
22 & EducationalInstitution & 1,799 & 3,031 & 862 \\
23 & PopulatedPlace & 1,743 & 9,081 & 694 \\
24 & School & 1,137 & 1,426 & 747 \\
\bottomrule
\end{tabular}
\end{table}

\subsubsection*{Summary}
We have shown that different entity types exhibit substantially different emergence patterns, but entities that belong to a particular type show broadly similar emergence patterns. 
Furthermore, entities with a fast transition from communicative to cultural \ac{CM}, are more likely to remain popular over time.


\section*{Conclusion}
\label{sec:discussion}
\label{sec:conclusion}

In this paper we studied entities as they transition from communicative into cultural \acl{CM}. We did so by studying a large set of time series of mentions of entities in online news streams before transitioning into cultural \ac{CM} (as represented by the creation of a Wikipedia page). 
We studied implicit groups of similarly emerging entities by applying a burst-based agglomerative hierarchical clustering method and explicit groups by isolating entities by whether they emerge in news or social media streams. 

\subsection*{Findings}
We found that, globally, entities have a long time span between surfacing in communicative \ac{CM} and transitioning into cultural \ac{CM}. 
During this time span, an entity may emerge with multiple bursts, however both the entities' introduction into public discourse, and subsequent transitioning into cultural \ac{CM} occur in the largest document bursts. Emergence statistics show large standard deviations, indicating that they differ substantially between entities. 
For this reason, we turned to time series clustering to uncover distinct groups of entities.
We discovered two emergence patterns: \acf{EB} entities and \acf{LB} entities. 
Analysis suggests that \ac{EB} entities comprise mostly ``head'' or popular entities; they exhibit fewer and higher bursts, with shorter emergence durations and lower emergence volumes. 
The \ac{LB} entities emerge more slowly, and witness a more gradual increase of exposure before transitioning into cultural \ac{CM}. 
The emergence patterns we visualized differ substantially from the global average and from, e.g., the type signatures shown in Figure~\ref{fig:classplots}, suggesting that the entities in each of the underlying clusters exhibit substantially different and distinct emergence patterns from entities in the other clusters. 

We showed that entities emerging in news and social media streams display very similar emergence patterns, but that on average, entities that emerge in social media take longer to be incorporated into cultural \ac{CM}. 
We hypothesize that this can be attributed to the nature of the underlying sources. 
News media are more mainstream and professional, with a larger audience, reach, and authority, than social media. 
Our findings are in line with those of \citet{enlighten82566}, who compare breaking news on traditional media with that on social media. 
Their findings suggest reported events overlap largely between both media, however, social media exhibits a long tail of minor events, which may explain the longer uptake on average. 
\citet{Leskovec:2009:MDN:1557019.1557077} find that the ``attention span'' for news events on social media both increases and decays at a slower rate than for traditional news sources, which additionally supports our observations of the slower uptake on social media. 

Finally, we showed that different entity types exhibit substantially different patterns, but entities of similar types show similar patterns. 
Some entity types, e.g., devices or creative works, transition faster from communicative to cultural \ac{CM}, than entities such as buildings, locations, and people. 
At the same time, the former ``fast'' entity types remain more popular over time. 

One aspect that distinguishes between ``fast'' and ``slow'' entity types, is that the former are more likely to appear in so-called ``soft news'' that covers sensational or human-interest events and topics (e.g., news related to celebrities and cultural artifacts). 
The slower entity types on the other hand, are more likely to appear in more substantive ``hard news'' that encompasses more urgent events and topics (e.g., political elections)~\cite[]{10.2307/2776752}. 
\citet{36915} studied the differences in ``attention span'' of the public 
and the traditional news media 
for ``hard'' and ``soft news,'' 
and found that hard news is associated with a relatively shorter period of public attention. 
Soft news exhibits a slower decrease of the public's attention, which supports our finding that faster entity types (more likely associated with soft news) tend to remain more popular over time. 

As emerging entities are not ``born equal.'' 
The patterns and circumstances under which an entity transitions from communicative to cultural \ac{CM} differ depending on source and type. 

\subsection*{Implications}
Our findings have implications for designing systems to detect emerging entities, and more generally for studying and understanding how collective memories are formed. 
We show that entities are likely to resurface multiple times in public discourse before transitioning into cultural \ac{CM}. 
This suggests that monitoring bursts of new entities could prove effective for predicting the formation of collective memories. 
Furthermore, we show that the type of stream in which entities emerge shows different patterns. 
This suggests that taking the different nature of streams into account can be beneficial for predicting emerging entities. 
Finally, we show that different types of entity exhibit different emergence patterns, suggesting the underlying entity type could likewise prove valuable in predicting emerging entities.

\subsection*{Limitations}
Part of our findings are derived from an unsupervised clustering method. 
Interpreting cluster signatures is a subjective matter, and clustering is a difficult task to evaluate~\cite[]{von2012clustering}.
In our defense, the clustering's dendrogram suggests the presence of distinct and meaningful groups, as the structure of the dendrogram shows symmetry and clear separations. 
More importantly, the cluster signatures yielded visually discernible, and different patterns between clusters, which was not the case for the signatures of other grouping strategies in this paper (see, e.g., Figure~\ref{fig:classplots}).

The fragmented nature of the source of our dataset (TREC-KBA StreamCorpus 2014) means that coverage, and hence representativeness of the data cannot be guaranteed. 
Popular social media channels such as Tumblr, Twitter and Facebook are not part of the dataset, there may be a sampling bias in the sources that represent the streams, resulting in a similar bias in the entities. 
Different sources may well yield different findings. 
This is unavoidable. 

Another limitation relates to the entity annotations used as a starting point in this paper: they cannot be assumed to be 100\% accurate. 
So-called ``cascading errors''~\citep{Finkel:2006:SPC:1610075.1610162} in NLP pipelines cause the accuracy of downstream tasks to suffer, in our case having imperfect tags (named entities) for imperfect tagging (entity linking). 
The FAKBA1 annotations are estimated (from manual inspection) to contain around 9\% incorrectly linked entities, with around 8\% of SERIF mentions being wrongfully not linked. 
Even more so, the ``difficult'' entities are long-tail entities, which are more likely to be part of our filtered set. 
However, manually correcting the annotations was beyond the scope of this study. 

Finally, there may be a cultural bias inherent in our choice of datasets: we used English language news sources and social media as well as the English version of Wikipedia. 
One could object that we studied the birth of collective memories for the \emph{English-speaking} part of the world, and that different datasets may also yield different findings. 
It is unfortunate that the English-speaking part of the world is disproportionately represented in our field of research, as is witnessed by the biggest constraint in conducting this study: dataset availability.  We invite the community to create suitable datasets in other languages or reflecting cultural practices in other parts of the planet so as to enable comparative studies.

\subsection*{Future work}
As a next step, we should take a closer look at the circumstances in which entities emerge, by not exclusively considering in how many documents they appear over time, but also in which contexts, e.g., by looking at the content of the articles themselves. 
Furthermore, in this paper we have chosen to restrict ourselves to the entities that transition and remain in cultural \ac{CM}. 
Another interesting aspect of \ac{CM}, is the notion of ``consensus.'' For example, one could study the emergence patterns of entities that are removed from cultural \ac{CM} after transitioning. 
Finally, the observations made in this paper could be explored in a prediction task, where, e.g., given a partial entity time series, the task would be to predict the point at which the entity transitions from communicative to cultural \ac{CM}.

\subsection*{Acknowledgements}
We thank our reviewers for their helpful comments that helped to improve the paper.

This research was supported by
Ahold Delhaize,
Amsterdam Data Science,
Blendle,
the Bloom\-berg Research Grant program,
the Dutch national program COMMIT,
Elsevier,
the European Community's Seventh Framework Programme (FP7/2007-2013) under
grant agreement nr 312827 (VOX-Pol),
the Microsoft Research Ph.D.\ program,
the Netherlands Institute for Sound and Vision,
the Netherlands Organisation for Scientific Research (NWO)
under pro\-ject nrs
727.\-011.\-005, 
612.001.116, 
HOR-11-10, 
CI-14-25, 
652.\-002.\-001, 
612.\-001.\-551, 
652.\-001.\-003, 
314-98-071, 
the Yahoo!\ Faculty Research and Engagement Program,
and
Yandex.
All content represents the opinion of the authors, which is not necessarily shared or endorsed by their respective employers and/or sponsors.

\bibliographystyle{apacite}
\setlength{\bibsep}{0pt}
\bibliography{jasist2016-collective-memories}

\end{document}